\documentclass[aps, prl, twocolumn,showkeys, showpacs,amsmath,amssymb,superscriptaddress]{revtex4-1}
\usepackage{graphicx}
\usepackage{dcolumn}
\usepackage{bm}
\usepackage{epsfig}
\usepackage{subfigure}
\usepackage{color}
\usepackage{epstopdf}

\begin{document}
\title{Relativistic effects on the Richtmyer-Meshkov instability}

\author{F. Mohseni} \email{mohsenif@ethz.ch} \affiliation{ ETH
  Z\"urich, Computational Physics for Engineering Materials, Institute
  for Building Materials, Wolfgang-Pauli-Strasse 27, HIT, CH-8093 Z\"urich
  (Switzerland)}

\author{M. Mendoza} \email{mmendoza@ethz.ch} \affiliation{ ETH
  Z\"urich, Computational Physics for Engineering Materials, Institute
  for Building Materials, Wolfgang-Pauli-Strasse 27, HIT, CH-8093 Z\"urich
  (Switzerland)}

\author{S. Succi} \email{succi@iac.cnr.it} \affiliation{Istituto per
  le Applicazioni del Calcolo C.N.R., Via dei Taurini, 19 00185, Rome
  (Italy),\\and Freiburg Institute for Advanced Studies,
  Albertstrasse, 19, D-79104, Freiburg, (Germany)}

\author{H. J. Herrmann}\email{hjherrmann@ethz.ch} \affiliation{ ETH
  Z\"urich, Computational Physics for Engineering Materials, Institute
  for Building Materials, Wolfgang-Pauli-Strasse 27, HIT, CH-8093 Z\"urich
  (Switzerland)} \affiliation{Departamento de F\'isica, Universidade
  Federal do Cear\'a, Campus do Pici, 60455-760 Fortaleza, Cear\'a,
  (Brazil)}

\date{\today}
\begin{abstract}
  Theoretical and numerical analysis of the relativistic effects on
  the Richtmyer-Meshkov (RM) instability reveals new and potentially
  very useful effects.  We find that, in contrast with the
  non-relativistic case, the growth rate of the RM instability depends
  strongly on the equation of state of the fluid, opening up the
  possibility to infer equations of state from experimental
  observations of the RM instability.  As opposed to the
  non-relativistic case, we also discover that, above a critical value
  of the fluid velocity, the growth rate of the instability
  counter-intuitively decreases due to the Lorentz's factor, and
  vanishes in the ultrarelativistic limit, as the speed of the
  particles approaches the speed of light.  Both effects might prove
  very useful for leading-edge applications, such as the study of the
  equation of state of quark-gluon matter, and the design of fast
  ignition inertial confinement fusion (ICF) schemes.  We perform a
  linear stability analysis to characterize the instability, for an
  arbitrary equation of state, and implement numerical simulations to
  study the instability in the non-linear regime, using the equation
  of state of an ideal gas. Furthermore, based on the numerical
  results, we propose a general expression that characterizes the long
  term evolution of the instability.
 \end{abstract}


\maketitle
The Richtmyer-Meshkov instability is one of the fundamental fluid
instabilities, which occurs when a shock wave passes through an
interface, separating two fluids with different densities.  This
instability was theoretically predicted by Richtmyer
\cite{richtmyer1960taylor} and experimentally detected by Meshkov
\cite{meshkov1969instability} in the non-relativistic context.  The
study of the RM instability is of major importance in several fields,
ranging from high energy physics \cite{goncharov1999theory} to
astrophysics \cite{arnett2000role}, especially wherever shock-wave
propagation is involved.  In a collapsing core supernova explosion,
the generated shock wave propagates outwards through a hydrogen-helium
interface. Observations have shown that the outer regions of the
supernovae are more uniformly mixed than expected, as a consequence of
the RM instability \cite{arnett2000role}. Therefore, the study of the
relativistic effects, from the theoretical point of view, can
contribute to a deeper understanding of this phenomenon.  On a very
different front, it has been recently discovered that quark-gluon
plasma (QGP) behaves as a nearly perfect fluid
\cite{PhysRevLett.91.182301, PhysRevLett.92.052302,
  PhysRevC.72.051901}, where shock waves have been theoretically
predicted \cite{PhysRevLett.32.741} and experimentally observed
\cite{0034-4885-52-10-003,PhysRevC.42.640}.  However, the equation of
state of this extreme state of matter is still under debate
\cite{fogacca2010nonlinear,
  meisinger2002phenomenological,begun2011modified}, and therefore,
developing new strategies to determine its form represents a very
important subject of research \cite{bouras2009relativistic}.  The
existence of relativistic shock-waves can lead to the appearance of
the RM instability.  In this Letter, we show that the RM instability
can also be used for this purpose, since, in contrast to the
non-relativistic case, the growth rate of the instability depends
explicitly on the equation of state.  As a result, experimental
information from the RM instability can be used to distinguish between
different theoretical models \cite{romatschke2012relativistic,
  huovinen2005anisotropy}.  The RM instability also plays a
significant role in inertial confinement fusion (ICF)
\cite{goncharov1999theory}, which has recently captured significant
attention as an alternative source of energy
\cite{hand2012laser}. Like most instabilities, the RM and
Rayleigh-Taylor instabilities represent one major cause of performance
degradation in energy applications \cite{wilson2004multifluid,
  ottaviani}.
\begin{figure}
  \centering \subfigure [Relativistic] {
\includegraphics[trim=35mm 20mm 15mm 5mm, clip, width=0.45\columnwidth, height=0.45\columnwidth ]{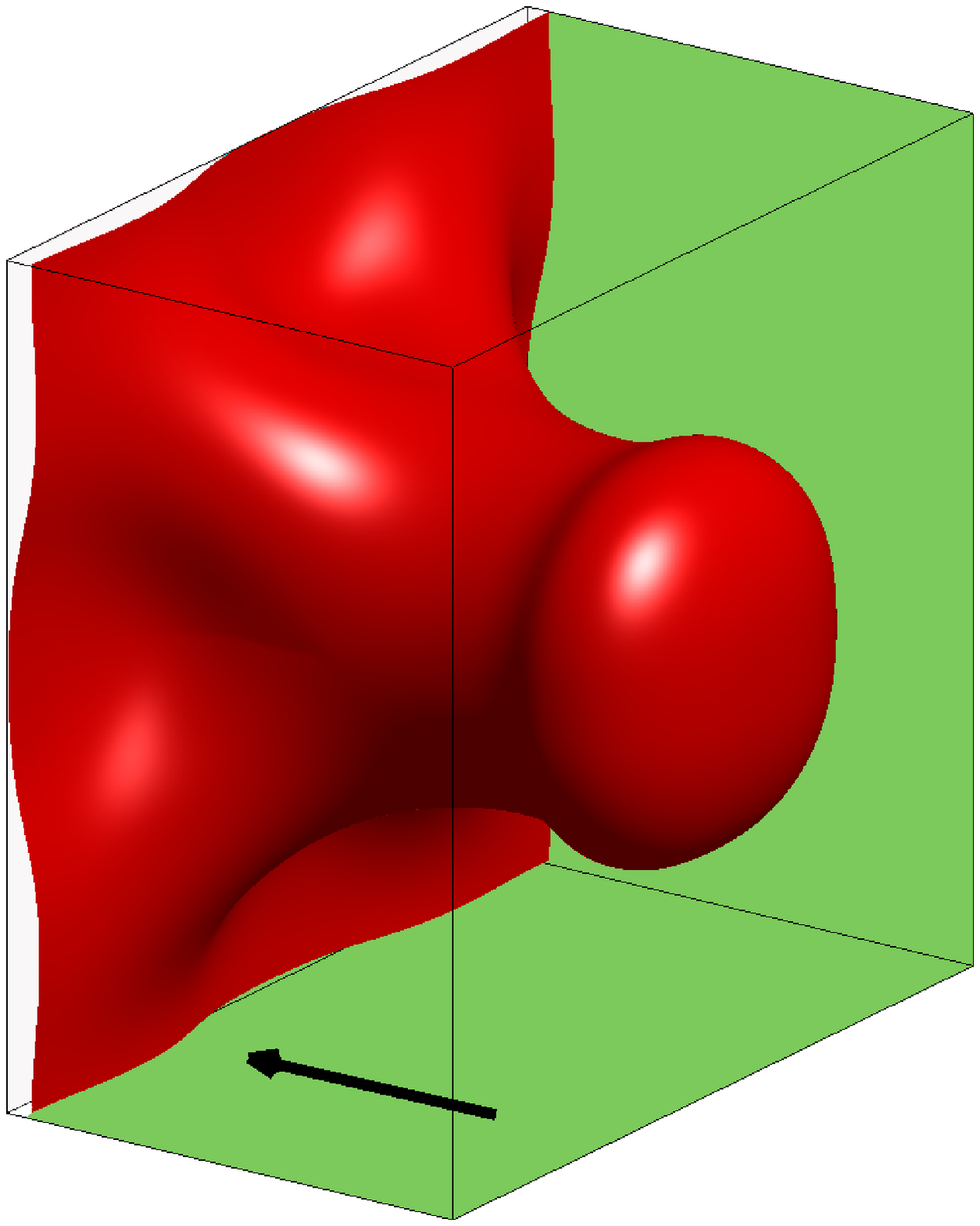}
\label{3Drel}
}
\subfigure [Non-Relativistic] {
\includegraphics[trim=35mm 20mm 15mm 5mm, clip, width=0.48\columnwidth, height=0.48\columnwidth ]{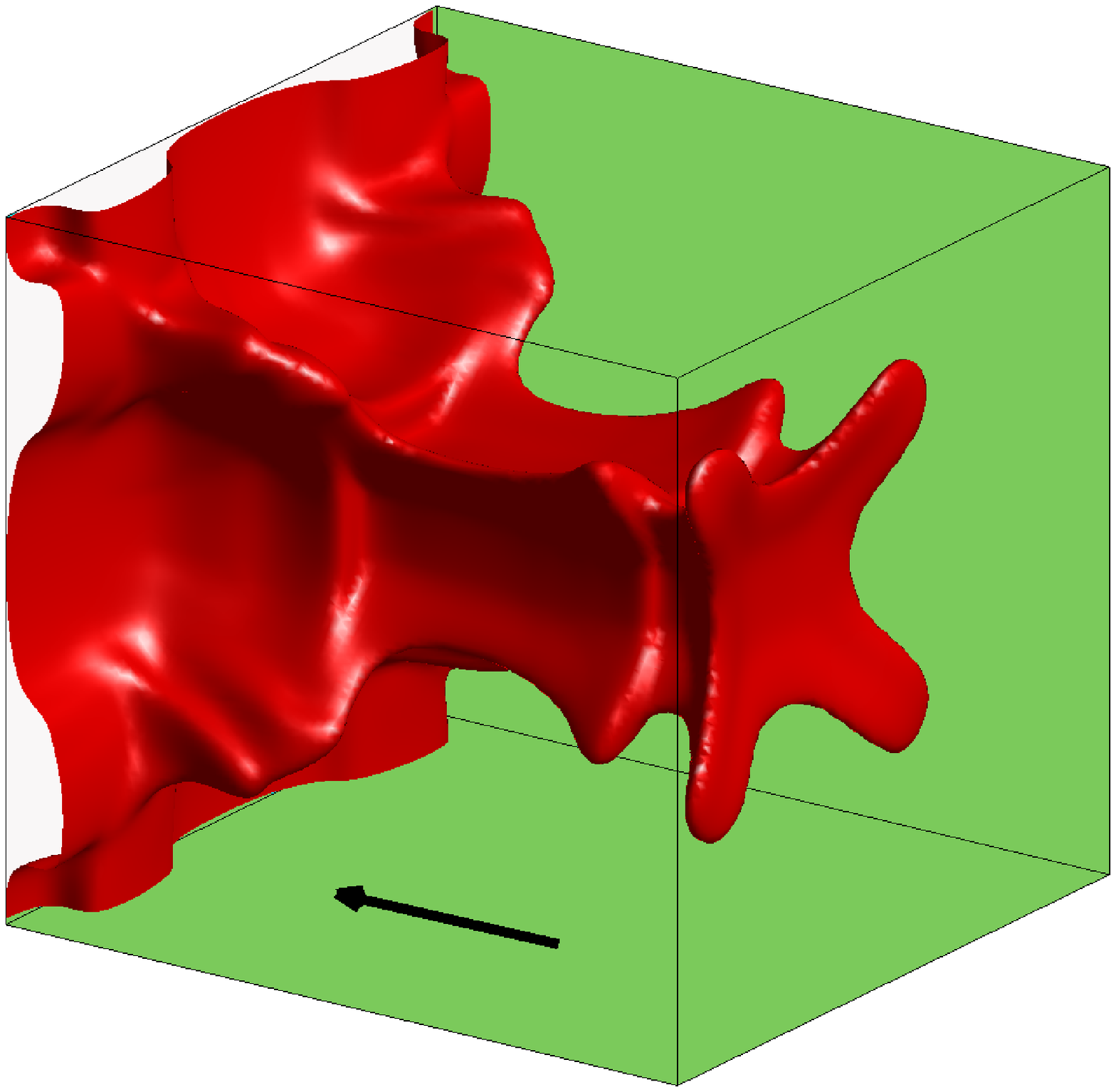}
\label{3Dclass}
}
\caption{Snapshots of the interface in the three dimensional shock tube
  Richtmyer-Meshkov instability at time $t=570$. The pre-shock density
  ratio is $28$ and the Mach number is $2.4$. Arrows show the
  direction of the shock wave.}
\label{3Dcomp}
\end{figure}
Fast ignition ICF approach uses a relativistic beam of electrons to
heat the compressed fuel, where relativistic fluid dynamic models can
describe the plasma-electron interactions
\cite{atzeni2005fluid,atzeni2009stopping}. Since there is not any
theoretical description for the RM instability in the relativistic
context, the design of better schemes becomes more difficult. Thus,
the results in this letter can be used to enhance the performance of
the fast ignition ICF schemes (see Fig.\ref{3Dcomp} for a comparison
between the relativistic and non-relativistic RM instabilities, where
we can appreciate the damping of the instability due to the
relativistic effects), as for instance, the fact that above a critical
value of the fluid velocity the RM instability presents a damping due
to the Lorentz's factor.

In the non-relativistic context, the RM instability has been
investigated extensively \cite{richtmyer1960taylor,wouchuk1996linear,
  wouchuk2001growth,richtmyer1960taylor,vandenboomgaerde1998impulsive,kotelnikov2000vortex}.
However, to the best of our knowledge, a systematic study of the RM
instability in the relativistic regime is still lacking.  In this
Letter, we first perform a linear stability analysis of the
relativistic RM instability and derive a theoretical asymptotic
expression for the growth rate of the perturbation amplitude. In order
to verify the theoretical results and to obtain a general expression
for the amplitude in the non-linear regime, we perform numerical
simulations of the relativistic RM instability.  To this purpose, the
recently introduced relativistic lattice Boltzmann model for high
velocities \cite{PhysRevD.87.083003, mendoza2010fast, hupp, karlin}
was extended to deal with the ideal gas equation of state. In order to
single out the relativistic effects, we also provide comparisons
between the relativistic and non-relativistic cases, and present a
study of the growth rate of the instability for different equations of
state.

Let us start by considering the growth of irregularities - in
particular sinusoidal corrugations - at the interface between two
fluids in the relativistic RM instability, i.e., when a relativistic
shock wave passes through the interface. In analogy to the
non-relativistic case \cite{richtmyer1960taylor}, we first approach
the problem by studying the Rayleigh-Taylor instability, which takes
place at the interface between two fluids at different densities,
whenever one of the two fluids accelerates into the other, and later
we replace the constant acceleration by an impulsive one representing
the shock wave.

The conservation equations of relativistic fluid dynamics are
$\partial_\alpha T^{\alpha \beta} = G^\beta$, and $\partial_\alpha
N^\alpha = 0$, where $T^{\alpha \beta}$ is the energy-momentum tensor,
$N^\alpha$ is the particle four-flow and $G^\beta$ is the force
density. For an ideal (inviscid) fluid we have $T^{\alpha
  \beta}=(\epsilon+p) U^\alpha U^\beta/c^2 - p \eta^{\alpha\beta}$,
and $N^\alpha=n U^\alpha$, where $p$ is the hydrostatic pressure,
$\epsilon$ the energy density (including the rest mass energy), $c$ is
the speed of light and $\eta^{\alpha\beta}$ is the Minkowski metric
tensor with the signature $(+,-,-,-)$. The macroscopic four-velocity
is $(U^\mu) = (c, \vec{u}) \gamma (u)$, with $\vec{u}$ being the
three-dimensional velocity and $\gamma(u)=1/\sqrt{1-u^2/c^2}$ the
Lorentz's factor. The relativistic force density can be defined as
$G^\alpha \equiv (\vec {F}.\vec{u} \gamma(u)/c, \vec{F} \gamma(u))$,
where $\vec{F}$ is the three-dimensional force density vector
\cite{cercignani}, and $n$ is the number of particles density.  Note
that the Einstein summation convention is assumed here and throughout
this paper. For the sake of simplicity, natural units i.e.,
$c=k_B=m=1$ are assumed hereafter. For the Rayleigh-Taylor
instability, we consider $\vec{F}=(\epsilon+p)\vec{g}$, where
$\vec{g}$ is the acceleration.

To calculate the amplitude growth rate of the disturbance at the
interface, we perform a linear stability analysis and without loss of
generality, we deal with this problem in two dimensions. Thus, small
perturbations are assumed for the velocity along $x$ and $y$
directions, i.e., $\delta u$ and $\delta v$, and the physical
variables, such as the density and pressure, i.e., $\delta p$ and
$\delta n$. For a single mode disturbance, we write
$A(x,y,t)=A_k(x)\exp(iky+\omega t)$, where $A$ stands for $\delta u$,
$\delta v$, $\delta p$ and $\delta n$, as well as the amplitude of the
perturbation $h$. Here $k=2\pi/\lambda$ is the wave number, $\lambda$
is the initial perturbation wavelength and $\omega$ is the wave
frequency of the perturbations.

We suppose that at $t=0$ the interface is located at $x=0$, and the
only non-zero component of $\vec{g}$ is in $x$ direction, i.e.,
$g$. To include the condition of incompressibility, and considering
the fact that second order terms play no role in the linear stability
analysis, the continuity equation simplifies to $\vec{\nabla} \cdot
\vec{u}=0$. Assuming that pressure and density are functions of $x$
only, we substitute the perturbed quantities in the conservation
equations and the incompressibility condition. Dropping the nonlinear
terms and considering initial equilibrium at the interface, i.e.,
$\partial p/\partial t=0$, we obtain a system of linear differential
equations.

Solving these equations, and taking $g/k \ll 1$, we find the following
dispersion relation, $\omega^2=(n_2-n_1) g k/(2
p+\epsilon_1+\epsilon_2)\gamma$, where $\epsilon_1$ and $\epsilon_2$
are the energy densities at both sides of the interface. In the
non-relativistic limit $\gamma \sim 1$, using the equation of state of
an ideal gas \cite{ryu2006equation}, i.e.,
$\epsilon+p=(\frac{1}{\Gamma -1}+1)p+n$, and considering $k_BT\ll
mc^2$ (so that pressure can be ignored with respect to the density),
we get the well known dispersion relation of the non-relativistic
Rayleigh-Taylor instability. Here $\Gamma$ is the adiabatic index,
i.e., specific heat at constant pressure divided by specific heat at
constant volume (see Supplementary Material \cite{supp}). Moreover, by
using the equation of state of an ideal gas with $\Gamma=4/3$ and
removing the Lorentz's factor, we obtain the relation for the
Rayleigh-Taylor instability for the ultra-relativistic case
\cite{allen1984rayleigh}. Hence, the amplitude of the perturbed
interface grows according to
\begin{equation} \label{secondpartial} 
\frac{\partial^2 h(t)}{\partial t^2}=\omega^2 h(t).
\end{equation}
In order to find a relation for the relativistic RM instability, we
replace the constant acceleration by an impulsive acceleration,
representing the shock wave. Let $\triangle u$ be the increment of
velocity due to this impulsive acceleration, we have $g(t)=\triangle u
\delta(t)$, where $\delta(t)$ is the Dirac delta function. Integrating
Eq.\eqref{secondpartial} and using the fact that $\int g(t) dt=
\triangle u$, we obtain the asymptotic relation for the growth rate of
the perturbation amplitude in the linear regime of the relativistic RM
instability:
\begin {equation}\label{firstpartial}
 v_f \equiv \frac{\partial h(t)}{\partial t} =\frac{(n_2-n_1) k h_0 \triangle u}{\gamma (2 p +\epsilon_2+\epsilon_1)}.
\end{equation}
Here, $h_0$ is the initial amplitude of the perturbation. Note that
this is a general expression which holds for any equation of the state
$\epsilon=\epsilon(T)$ and $p=p(T)$, where $T$ is the fluid
temperature.  This shows that, unlike its non-relativistic
counterpart, the growth rate of the amplitude for the relativistic RM
instability depends on the equation of state.

In the RM instability, the light fluid penetrates the heavy one,
generating bubbles and the heavy fluid penetrates the light one,
giving rise to spikes. The perturbation amplitude, $h(t)$, is
calculated by measuring the distance between the tips of the spike and
the bubble divided by two. Note that the linear assumption is well
justified only as long as the interface amplitude is small, i.e.,
$h/\lambda < 0.1$ \cite{brouillette2002richtmyer} and nonlinear
effects become important when the amplitude becomes larger.

One can immediately notice from Eq.\eqref{firstpartial} that
relativistic effects decrease the amplitude growth rate compared to
the non-relativistic RM instability, due to the Lorentz's factor,
$\gamma>1$, as well as to the contribution of the pressure to the
inertia of the relativistic fluid, which becomes relevant at high
temperatures. This argument is in line with previous observations in
\cite{mendoza2011preturbulent}, which show a delay in the onset of
pre-turbulence of relativistic electronic micro-jets in graphene.

Note that, in the presence of the shock wave, Eq.\eqref{firstpartial}
permits to find the asymptotic linear growth rate of the amplitude,
using the post-shock values of $n_1$, $n_2$ and $h_0$, $\triangle u$
and $p$ being the velocity jump and pressure at the interface.

To find a general relation for the amplitude in the nonlinear regime,
we perform numerical investigations of the relativistic RM
instability, using the lattice Boltzmann model for high velocities
recently proposed in \cite{PhysRevD.87.083003} (for more details see
Supplementary Material \cite{supp}). Note that using the relativistic
Boltzmann model leads to viscous hydrodynamics. However, it has been
shown that viscosity has a negligible effect on the perturbation
amplitude in the non-relativistic shock tube RM instability
\cite{carles2002effect, jones1997membraneless}, so that we expect a
similar behavior in relativistic hydrodynamics.
\begin{figure}
  \centering \subfigure [Relativistic] {
\includegraphics[trim=57mm 70mm 30mm 75mm, clip, width=1.0\columnwidth, height=0.4\columnwidth ]{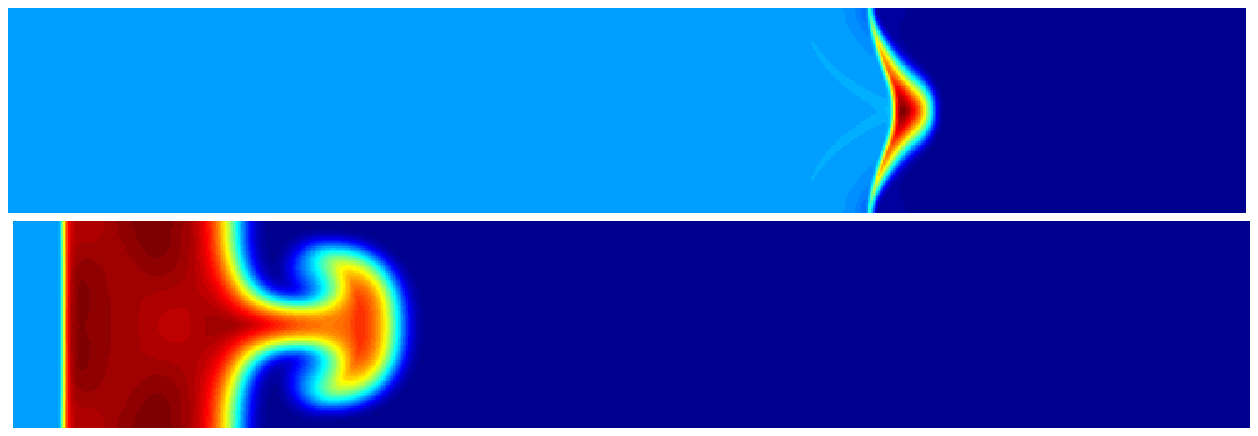}
\label{densityrel}
} \subfigure [Non-Relativistic] {
\includegraphics[trim=50mm 75mm 31mm 67mm, clip, width=1.0\columnwidth, height=0.4\columnwidth ]{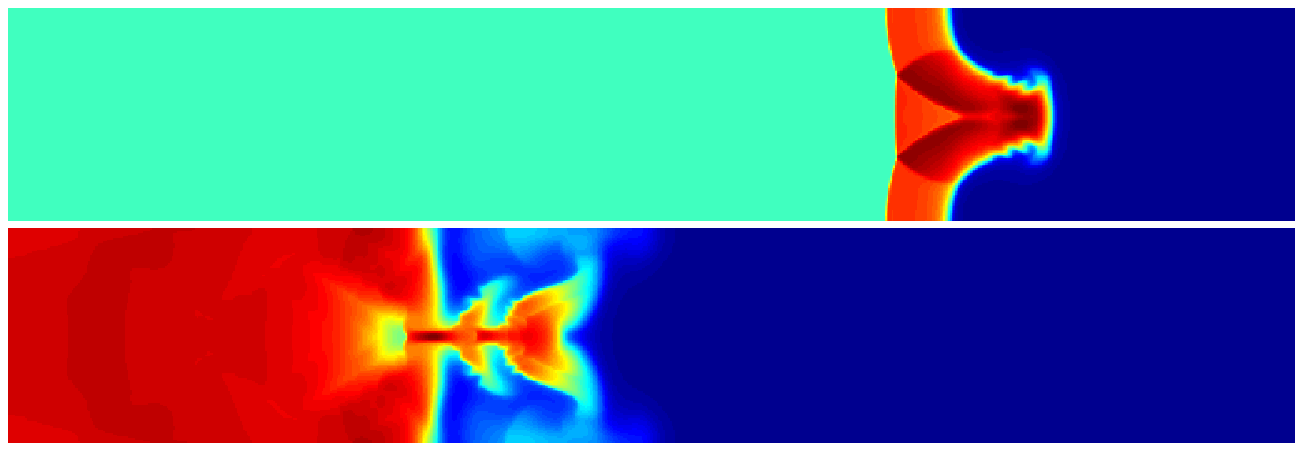}
\label{densityclass}
}
\caption{Snapshots of the density field in the two dimensional shock
  tube Richtmyer-Meshkov instability for (a) relativistic and (b)
  non-relativistic cases at different times. For both cases, from top
  to bottom, snapshots corresponds to the times $t=180$ and $t=1260$,
  respectively. Here, $n_L/n_M=28$ and Mach number is 2.4. Blue and
  red colors denote low and high densities, respectively. Here, time
  represents the number of time steps multiplied by $\delta t/\delta x
  = 0.15$. }
\label{densitytime}
\end{figure}
For the two-dimensional simulation of the shock tube RM instability, a
domain of $200\times1200$ lattice cells is considered.  Since we use
dimensionless numbers to characterize the RM instability, we refer to
numerical units throughout this Letter. For all the simulations
considered here, a shock wave with the velocity
$\beta=|\vec{u}|/c=0.94$, traveling from right to left, is passing
through a sinusoidal perturbation in the density located at $x_p=1000$
cells. The initial position of the shock wave is at $x_s=1100$
cells. The single mode sinusoidal perturbation at the interface is:
$x=x_p+a \sin(\frac{\pi}{2}+\frac{2\pi}{\lambda}y)$, where $a$ is the
pre-shock amplitude of the interface and $\lambda$ is the width of the
domain. Note that, hereafter the subscripts $R$, $M$, and $L$ refer to
the right hand side of the shock, the region between the shock and the
initial perturbation, and the left hand side of the perturbation,
respectively.  The densities at the two sides of the perturbation are
different, and the pressure is forced to be constant across the
perturbation, i.e., $p_M=p_R$, by choosing appropriate values of the
temperature. For simplicity, the simulations have been performed
assuming the equation of state of an ideal gas for various pre-shock
density ratios $n_L/n_M$ and various values of the relativistic Mach
number of the shock wave $Ma_r=u_s \gamma(u_s)/c_s \gamma(c_s)$, where
the velocity of the shock $u_s$ and the sound velocity $c_s$ are
defined as \cite{Rischke1995346,ryu2006equation}
\begin{equation}
  u_s^2=\frac{(p_R-p_M)(\epsilon_R+p_M)}{(\epsilon_M+p_R)(\epsilon_R-\epsilon_M)},
  \quad c_s^2=\frac{\Gamma (\Gamma-1) \frac{p_M}{n_M}}{\Gamma \frac{p_M}{n_M}+\Gamma-1}.
\end{equation}
The values of initial densities and pressures are calculated in such a
way as to obtain the desired values of $n_L/n_M$ and $Ma_r$, while the
velocity of the shock is fixed. For more details, see Supplementary
Material \cite{supp}.

Fig.~\ref{densityrel} shows the density field and the evolution of the
bubble and the spike for the case $n_L/n_M=28$, $Ma_r=2.4$ and $a=32$
at different times, after the shock wave has passed through the
initial perturbation. Finally, the spike forms the characteristic
mushroom shape of the instability. It is worth mentioning that the
passage of the shock wave through the heavy fluid causes an increase
in its density, due to the compression, which is well visible in
Fig.~\ref{densitytime}.

For the purpose of comparison, we have also performed a numerical
simulation for the non-relativistic RM at the same density ratio and
Mach number as in Fig.~\ref{densityrel}, i.e., $n_L/n_M=28$ and
$Ma=2.4$. The results are presented in Fig.~\ref{densityclass}.  Here,
in order to draw an accurate comparison between the two cases, and
regarding the fact that we are simulating viscous hydrodynamics, the
Reynolds number should also be the same for both cases. Thus,
following Ref.\cite{PhysRevLett.103.025301}, we define the
relativistic Reynolds number for the shock tube relativistic RM
instability as $Re_r=(\epsilon+p) u_s \gamma(u_s) \lambda/\eta$, where
$\eta$ is the shear viscosity. For the non-relativistic numerical
simulations, we have used the model proposed in
Ref.\cite{li2007coupled}. Fig.~\ref{densityclass} shows that, in the
non-relativistic RM, the amplitude of the perturbation grows much
faster at early times, leading to a faster development and more
complex structures of the instability at later times. In fact, this
agrees with our analytical results, Eq.\eqref{firstpartial}, where we
argued that relativistic effects lead to a damping of the
instability. This can be also seen in Fig.~\ref{3Dcomp}, where the
results of the simulation for the 3D shock tube RM instability with
square cross section is presented, with $n_L/n_M=28$ and the
$Ma_r=Ma=2.4$. The 3D simulation was performed with the same
parameters as before and using a lattice size of
$200\times200\times1200$ cells.
\begin{figure}
\begin{center}
\includegraphics[trim=10mm 2mm 14mm 0mm, clip, width=0.9\columnwidth, height=0.8\columnwidth ]{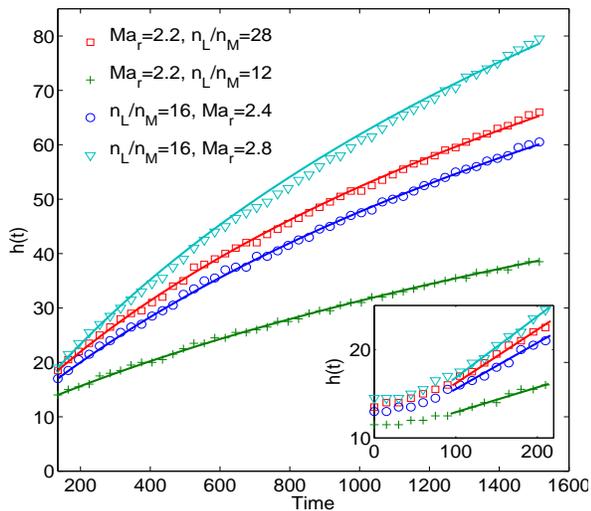}
\end{center}
\caption{Results of the numerical simulation $h(t)$ versus time for
  different $Ma_r$ and density ratios in the nonlinear regime. Solid
  lines are the resulting $h(t)$, using the proposed relation,
  Eq.\eqref{fitting}. In the inset, the results for linear regime are
  presented and the solid lines show the asymptotic theoretical growth
  rate, Eq.\eqref{firstpartial}.}
\label{htimemach}
\end{figure}
\begin{figure}
\begin{center}
\includegraphics[trim=0mm 5mm 5mm 0mm, clip, width=0.9\columnwidth, height=0.8\columnwidth ]{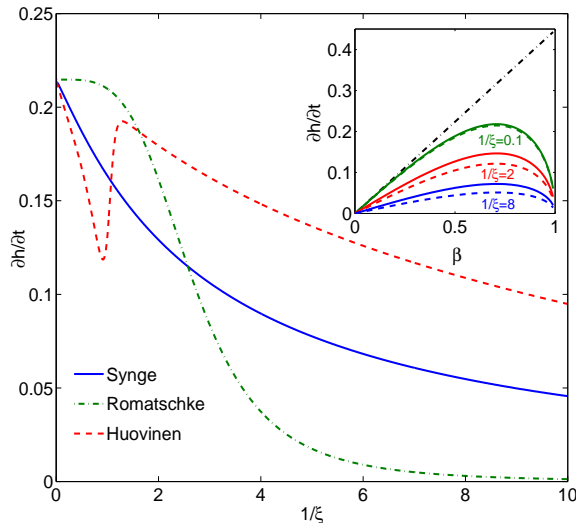}
\end{center}
\caption{Perturbation growth rate in the linear regime calculated from
  Eq.\eqref{firstpartial} as a function of temperature for different
  equations of state proposed in Refs. \cite{synge1957relativistic},
  \cite{romatschke2012relativistic} and
  \cite{huovinen2005anisotropy}. In the inset, we see the effect of
  increasing $\beta=\triangle u/c$ on the perturbation growth rate in
  the linear regime (using the ideal gas equation of state: solid
  lines for $\Gamma=5/3$ and dashed lines for $\Gamma=4/3$), for
  different temperatures, comparing with the non-relativistic RM
  instability (dashed-dotted line).}
\label{compclass}
\end{figure}

Since Eq.\eqref{firstpartial} predicts an asymptotic growth rate of
the amplitude in the linear regime, we are now interested in a general
relation for the growth rate of the amplitude of the instability in
the nonlinear regime. For the shock tube RM instability and for fixed
adiabatic index and wave number and in the absence of surface tension,
the amplitude depends on density ratio and Mach number. Therefore,
several simulations have been performed for different density ratios
and relativistic Mach numbers, i.e., $8\leq n_L/n_M \leq 28$ and $2
\leq Ma_r \leq 3$, when $a=16$. Fig.~\ref{htimemach} shows the
numerical results of $h(t)$ for different values of $Ma_r$ and density
ratios in the nonlinear regime. By increasing $Ma_r$, as well as the
density ratio, the amplitude grows faster.  Note that $v_f$ can be
used as the initial growth rate of the nonlinear regime. In order to
describe the effect of the nonlinearity on the growth rate, we propose
the following relation (see Fig.~\ref{htimemach}),
\begin{equation} \label{fitting} \frac{\partial h(t)}{\partial t}=
  \frac{v_f}{1+a_1 (\frac{n_L}{n_M})^{1/2} Ma_r t},
\end{equation}
where $a_1=9.033\times10^{-5}$ is a constant value. Note that this
constant is independent of the Mach number and the density ratio, and
therefore we can assume that this is a universal constant
characterizing the instability.  In the inset of Fig.~\ref{htimemach},
the results of $h(t)$ for the linear regime are presented, where the
slope of the solid lines represents the theoretical growth rate,
Eq.\eqref{firstpartial}.  As expected, at early times, the
compressibility effects decrease the growth rate, but at later times,
when the compressibility effects are weaker, very good agreement is
found between the theoretical, Eq.~\eqref{firstpartial} and numerical
growth rates.

In order to explore the consequences of the relativistic effects, we
plot the asymptotic linear growth rate of the instability,
Eq.\eqref{firstpartial} as a function of the temperature, for
different equations of state (see Fig.~\ref{compclass}). Here, we
compare the equation of state for an ideal relativistic gas developed
by Synge \cite{synge1957relativistic}, the non-ideal equation of state
for QGP proposed by Romatschke \cite{romatschke2012relativistic}, and
the equation of state for strongly interacting matter in relativistic
heavy ion collision (with a phase transition) proposed by Huovinen
\cite{huovinen2005anisotropy}. We have defined the dimensionless
parameter $\xi=mc^2/k_BT$ and the following parameters are considered:
$h_0=20, k=0.025, \triangle u=0.6$, and the post-shock density ratio
is $18$. Fig.~\ref{compclass} shows that the perturbation growth rate
strongly depends on the equation of state, and therefore, the RM
instability can be used to check the validity of a proposed equation
of state. In the inset of Fig.~\ref{compclass}, the linear growth rate
of the amplitude versus the velocity, $\beta=\triangle u/c$, for the
ideal gas equation of state, is compared with the respective relation
for the non-relativistic RM instability, at different temperatures.
Note that for low velocities and low temperatures, both relations
agree, but by further increasing $\beta$ and/or the temperature, the
perturbation amplitude decreases as compared to the non-relativistic
case. Note that the value of $\beta=1/\sqrt 2$ optimizes the growth of
the perturbation in the relativistic RM instability and for the case
of massless particles, $\beta=1$, the growth rate of the instability
vanishes. Additionally, each relativistic result in the inset of
Fig.~\ref{compclass} is presented for two cases, $\Gamma=5/3$ and
$\Gamma=4/3$, and, as expected, the results with $\Gamma=5/3$ are
closer to the non-relativistic ones.

Summarizing, we have shown that the RM instability can be exploited to
distinguish between different equations of state, by measuring the
growth rate of the instability. This result can be used as a method to
determine the equation of state of QGP, which is still under
debate. In addition, by increasing the jump velocity across the
interface above a critical value, the growth of the instability
decreases and eventually goes to zero for $\beta=1$. This is an
unexpected result because one would expect that for higher velocities
the fluid should become more unstable. This effect can be exploited
for improving the design of fast ICF schemes, where damping of the
instability would prove beneficial to the overall efficiency of the
process. We have also proposed a long-term relation for the evolution
of the interface amplitude for different values of the density ratio
and relativistic Mach number.  To study analytically the relativistic
effects, we have developed a linear impulsive model, based on the
linear instability analysis, to predict the asymptotic amplitude
growth rate of the interface in the linear regime. Developing new
theoretical results for compressible cases and non-linear regimes, and
investigating other types of equation of state, makes a very
interesting subject for further research.

\begin{acknowledgments}
  We acknowledge financial support from the European Research Council
  (ERC) Advanced Grant 319968-FlowCCS. The authors are also grateful
  for the financial support of the Eidgenössische Technische
  Hochschule Z\"urich (ETHZ) under Grant No. 0611-1.
\end{acknowledgments}

\bibliography{referencesRM}

\end{document}